\documentclass[preprintnumbers]{revtex4}

\usepackage{epsfig}
\usepackage{graphicx}
\def\Ord{\buildrel{\scriptscriptstyle <}\over{\scriptscriptstyle\sim}}
\def\OOrd{\buildrel{\scriptscriptstyle >}\over{\scriptscriptstyle\sim}}
\def\ZZ {\hbox{\it Z\hskip -4.pt Z}}
\def\tanb{\tan\beta}
\def\wt{\widetilde}
\def\be{\begin{equation}}
\def\ee{\end{equation}}
\def\bea{\begin{eqnarray}}
\def\eea{\end{eqnarray}}
\def\nn {\nonumber}
\def\Jnl #1#2#3#4 {#1 {\bf #2}, (#3) #4}
\def\NPB {{\em Nucl. Phys.} {\bf B}}
\def\PLB {{\em Phys. Lett.}  {\bf B}}

\def\PRD {{\em Phys. Rev.} {\bf D}}
\def\ZPC {{\em Z. Phys.} {\bf C}}
\def\EPJC {{\em Eur. Phys. J.} {\bf C}}

\begin{document}

\preprint{~RAL-TR-2001-036}
\preprint{~CERN-TH/2001-241}
\preprint{~IPPP/01/46}

\title{Higgs Sector of Non-minimal Supersymmetric 
Models at Future Hadron Colliders}



\author{Cyril Hugonie}
\email[]{C.Hugonie@durham.ac.uk}
\affiliation{Institute for Particle Physics Phenomenology,
University of Durham, Durham DH1 3LE, UK}

\author{Stefano Moretti}
\email[]{Stefano.Moretti@cern.ch}
\affiliation{CERN~Theory~Division,~CH-1211~Geneva~23,~Switzerland\\
Institute~for~Particle~Physics~Phenomenology,~University~of~Durham,~Durham~DH1~3LE,~UK}


\date{\today}

\begin{abstract}
We investigate the potential of current and planned hadron colliders operating
at the TeV scale in disentangling the structure of the Higgs sector of
non-minimal Supersymmetric extensions of the Standard Model with an extra gauge
singlet. We assume universality of the soft Supersymmetry breaking terms at the
GUT scale as well as a CP-even Higgs boson with mass around 115 GeV, as
suggested by LEP. We find that mixing angles between the doublet and singlet
Higgs states are always small. However, 
concrete prospects exist at both the Tevatron
(Run II) and the Large Hadron Collider of detecting at least one neutral Higgs
state with a dominant singlet component, in addition to those available from
a doublet Higgs sector which is similar to 
the one of the Minimal Supersymmetric
Standard Model.
\end{abstract}

\maketitle


\vspace{-1.0truecm}
\section{Introduction}\label{intro}

The NMSSM \cite{NMSSM1,NMSSM2} (Next-to-Minimal 
Supersymmetric Standard Model) is
defined by the addition of a gauge singlet Superfield $S$ to the 
MSSM (Minimal Supersymmetric Standard Model) and a
global $\ZZ_3$ symmetry on the renormalizable part of the Superpotential. It
allows to omit the  term $\mu H_u H_d$ in the Superpotential of
the MSSM (where $H_u$ is the Higgs doublet coupled to the up-type fermions and 
$H_d$ to the down-type ones) and to replace it by a Yukawa coupling
(plus a singlet self coupling), hence solving the 
so-called `$\mu$ problem' of the MSSM.
Apart from the standard quark and lepton Yukawa couplings, the Superpotential
of the NMSSM is 
\be
W = \lambda H_u H_d S + {1\over3} \kappa S^3 + ... \label{supot}
\ee
and the corresponding trilinear couplings $A_\lambda$ and $A_\kappa$ are added
to the soft  Supersymmetry 
(SUSY) breaking terms. Once the electroweak (EW) symmetry is broken, the
scalar component of $S$ acquires a vacuum expectation value
(vev) $s = \langle S \rangle$, thus generating
an effective $\mu$ term, $\mu = \lambda s$. The Superpotential (\ref{supot}) is
scale invariant, and the EW scale appears only through the soft SUSY
breaking terms. The possible domain wall problem due to the spontaneous
breaking of the $\ZZ_3$ symmetry at the weak scale is assumed to be solved by
adding non-renormalizable interactions which break the $\ZZ_3$ symmetry without
spoiling the quantum stability with unwanted divergent singlet tadpoles
\cite{walls}. This can be done by replacing the $\ZZ_3$ symmetry by a discrete
$R$-symmetry, broken by the soft SUSY breaking terms \cite{Greeks}.

A similar model, called nMSSM (for new Minimal Supersymmetric Standard Model)
has recently been proposed \cite{nMSSM}, using discrete $R$-symmetries to
forbid the singlet self-interaction in (\ref{supot}). In
this model, $n$-th order singlet tadpole graphs generate a divergent
loop-suppressed term in the scalar potential,
$
V_{\mathrm{tadpole}} \sim \frac{1}{(16\pi^2)^n} M_{\mathrm{SUSY}}^2 M_P (S+S^*)
\equiv \xi^3 (S+S^*)$ ($\xi \sim M_{\mathrm{SUSY}}$). 
This term breaks the dangerous Peccei-Quinn
symmetry present when $\kappa$ is set to 0 in the NMSSM.

The new states in both models are one additional CP-even neutral Higgs $S_r$
(real part of the complex scalar field $S$), one CP-odd neutral Higgs $S_i$
(imaginary part), as well as one additional neutralino, the 
`singlino' $\wt{S}$.
These states usually mix with the corresponding MSSM states, giving three
CP-even neutral ones, two CP-odd neutral Higgses and five neutralinos.

In this study, we focus on the phenomenology of the neutral Higgs sector of the
NMSSM and nMSSM at the Tevatron (Run II, $\sqrt s=2$ TeV) and the Large Hadron
Collider (LHC, $\sqrt s=14$ TeV). We only consider the `direct' production
channels, namely \cite{direct}
($V=W^\pm,Z$, $Q=b,t$ and $q^{(')}$ refers to any possible quark
flavour):
\bea\label{procs}
gg \to \mathrm{Higgs} ~ ({\mathrm{gluon-gluon~fusion}}),
& &
q\bar q^{(')} \to V~\mathrm{Higgs} ~ ({\mathrm{Higgs-strahlung}}), \nn \\
q q^{(')}\to q q^{(')}~\mathrm{Higgs} ~ (VV-{\mathrm{fusion}}),
& &
gg,q\bar q\to Q\bar Q~\mathrm{Higgs} ~
({\mathrm{heavy-quark~associated~production}}).
\eea
Here, we neglect `indirect' Higgs production via 
decays/bremsstrahlung off heavy SUSY particles \cite{indirect}.

\section{Parameter space of the models}\label{setup}

In order to study the Higgs spectrum of both models, we have numerically
scanned their free parameter spaces using a similar program as described in
Refs.~\cite{NMSSM2,nMSSM}. First, we constrained the soft terms of both models
by requiring {\em universality} at the GUT scale. The independent parameters of
the models are then: a universal gaugino mass $M_{1/2}$, a universal mass for
the scalars $m_0$, a universal trilinear coupling $A_0$, the Yukawa coupling
$\lambda$ and the the singlet self-coupling $\kappa$ (NMSSM) or the tadpole
coefficient $\xi$ (nMSSM). The (well-known) value of the $Z$-boson mass fixes
one of these parameters with respect to the others, so that we end up with {\em
four} free parameters at the GUT scale, i.e., as many as in the MSSM with
universal soft terms. In principle, one could choose the same set of free
parameters as in the MSSM, i.e., $M_{1/2}$, $m_0$, $A_0$ and $\tanb (\equiv
\frac{h_u}{h_d})$, with $s$, $\lambda$, and $\kappa$ ($\xi$) being determined
by the three minimisation equations. However, this appears to be not easily 
feasible, as $\lambda$ ($\kappa$) also influences the running of the
renormalization group equations (RGEs) for the soft parameters between the GUT
and the weak scale. Therefore, we took the following input parameters:
$m_0/M_{1/2}$, $A_0/M_{1/2}$, $\lambda$ and $\kappa$ ($\xi/M_{1/2}$). We then
integrated numerically the RGEs between the GUT and the weak scale and
minimised the two-loop effective potential. This gives $\tanb$ and $s$ and the
overall scale $M_{1/2}$ is fixed by $M_Z$. Finally, we imposed the current
experimental bounds on (s)particle masses and couplings, especially the 
LEP limits
on the Higgs mass vs. its coupling to gauge bosons, see
\cite{LEP1}. Furthermore, we assumed the existence of one neutral CP-even Higgs
boson with mass 115 GeV and sufficient coupling to gauge bosons, as suggested
by LEP \cite{LEP2}.

\section{Results}\label{results}

The main result of this numerical analysis, as already pointed out in
Refs.~\cite{NMSSM2,nMSSM}, is that the additional couplings appearing in 
 (\ref{supot}) are always small: $\lambda (\kappa) < 10^{-2}$ 
(NMSSM) and $\lambda<0.2$ (nMSSM). (Higher values would lead to unphysical
minima of the scalar potential.) The mixing angles of the additional singlet
states (described in Sect. \ref{intro}) to the non-singlet sector, being
proportional to these couplings, are always small and the singlet sector of the
universal NMSSM/nMSSM is quasi decoupled. (In the non-universal scenario, the
outcome may be quite different: see Ref.~\cite{NonUni}). Hence, the neutral
Higgs sector consists of a quasi pure (qp) CP-even Higgs singlet state, $S_r$,
a qp CP-odd singlet, $S_i$, and the doublet sector is basically MSSM-like,
apart from small perturbations of order $\sim \lambda^2$, so that results known
for the Higgs sector of the MSSM are also valid in our case.

Fixing the mass of the lightest visible (non-singlet) CP-even Higgs at 115 GeV
puts further constraints on the parameter space of both models: we find that
$\tan\beta$ is always larger than 4, the CP-odd doublet Higgs mass $M_A$ is
larger than 160 GeV and $M_{\mathrm{SUSY}}$ is larger than 350 GeV. In this
limit, the CP-even  doublet states are qp interaction eigenstates. The Higgs
state with mass 115 GeV is a qp $H_u$, and the qp $H_d$ is heavy (with mass
larger than 300 GeV). On the other hand, the masses of the singlet Higgs
states, $S_r$ and $S_i$, can vary from a few GeV to 1 TeV.

\begin{figure}[t]
\begin{minipage}[b]{.495\linewidth}
\hskip1.25cm{\small Total number of $S_r$ events, $N_{S_r}$}
\centering\epsfig{file=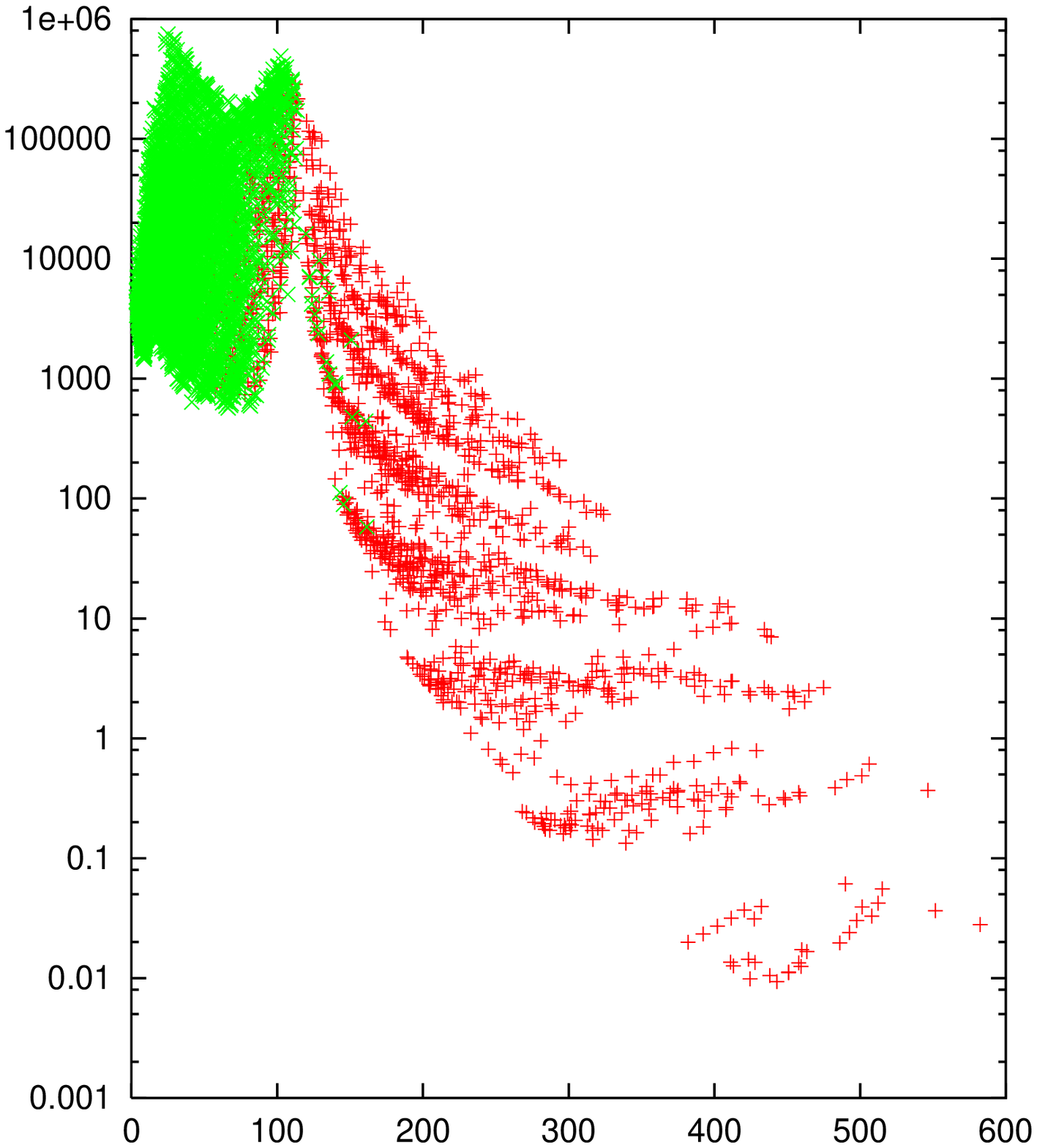,angle=0,height=5cm,width=\linewidth}
\end{minipage}\hfill\hfill
\begin{minipage}[b]{.495\linewidth}
\hskip1.25cm{\small Total number of $A$ events, $N_A$}
\centering\epsfig{file=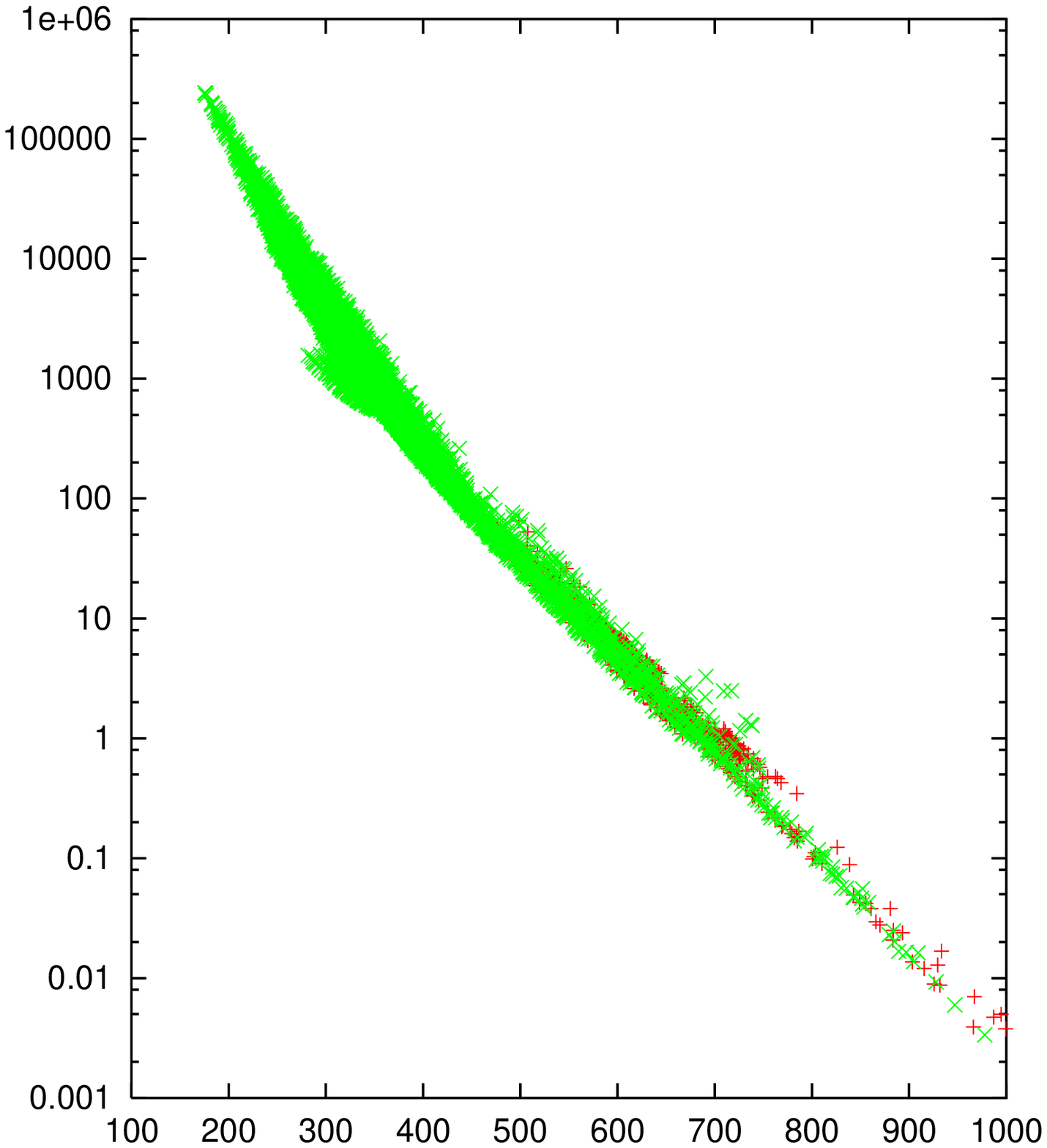,angle=0,height=5cm,width=\linewidth}
\end{minipage}\hfill\hfill
\begin{minipage}[b]{.495\linewidth}
\centering\epsfig{file=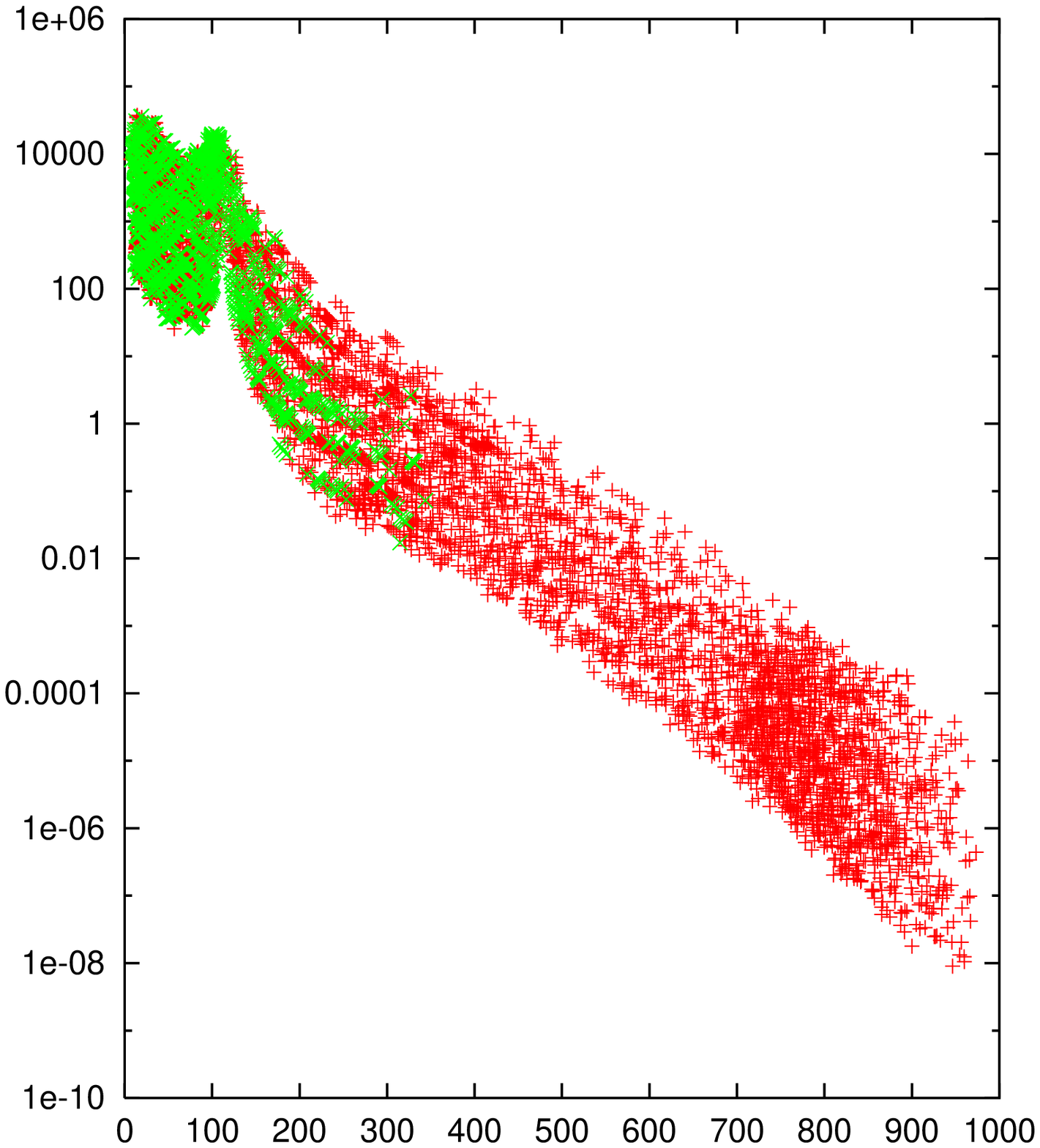,angle=0,height=5cm,width=\linewidth}
\end{minipage}\hfill\hfill
\begin{minipage}[b]{.495\linewidth}
\centering\epsfig{file=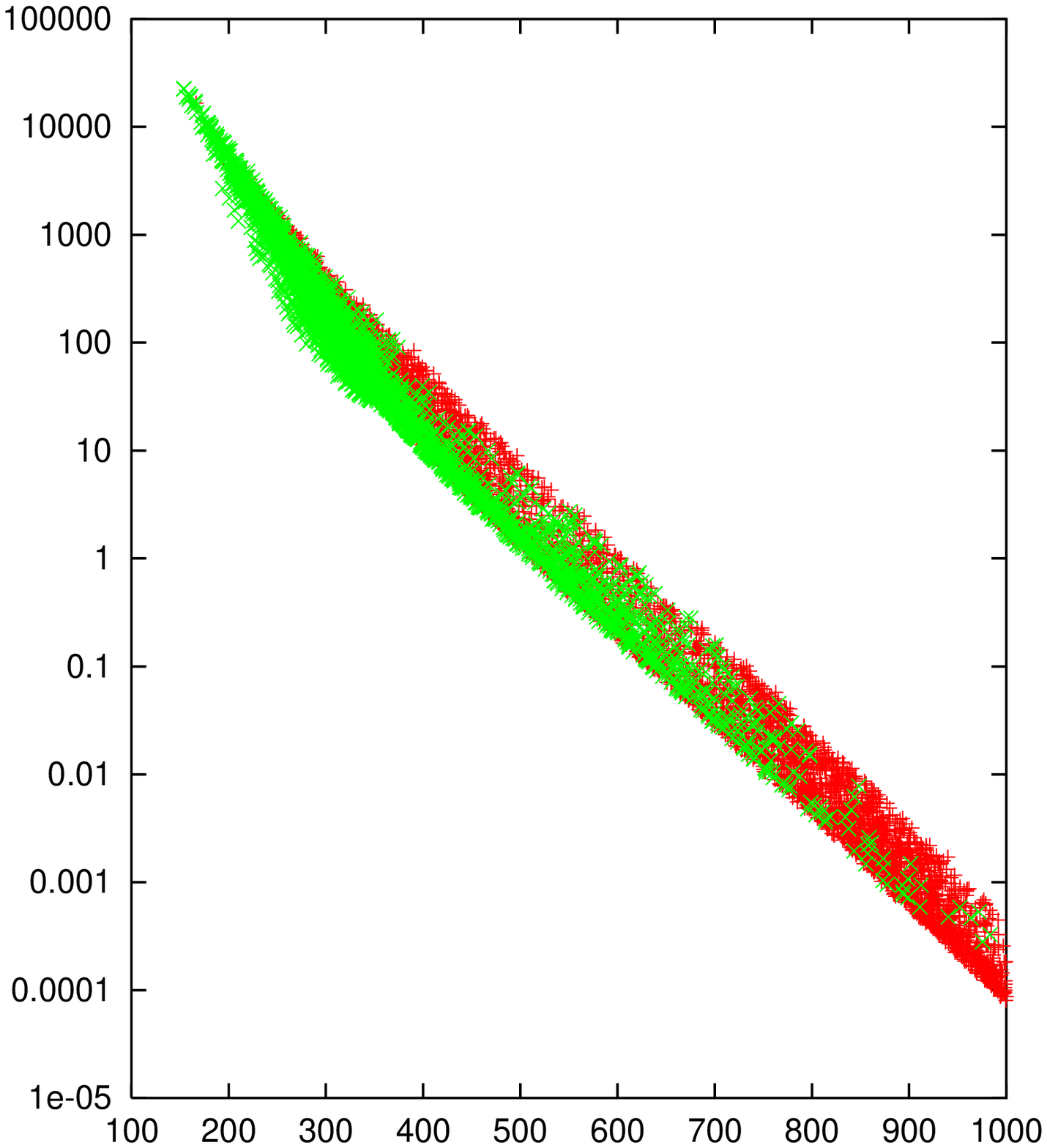,angle=0,height=5cm,width=\linewidth}
\end{minipage}

\hskip4.0cm{\small $M_{S_r}$ (GeV)}\hskip6.75cm{\small $M_A$ (GeV)}
\caption{Total number of events produced through processes (\ref{procs}) at the
Tevatron (Run II) after 15 fb$^{-1}$ in the NMSSM (upper plots) and nMSSM (lower
plots) for the CP-even singlet $S_r$ (left plots) and the CP-odd non-singlet $A$
(right plots) as a function of the produced Higgs mass. (For an explanation of
the colour coding, see the text).}
\vspace*{-3mm}
\label{fig:Tevatron}
\end{figure}

\begin{figure}[t]
\begin{minipage}[b]{.495\linewidth}
\hskip1.25cm{\small Total number of $S_r$ events, $N_{S_r}$}
\centering\epsfig{file=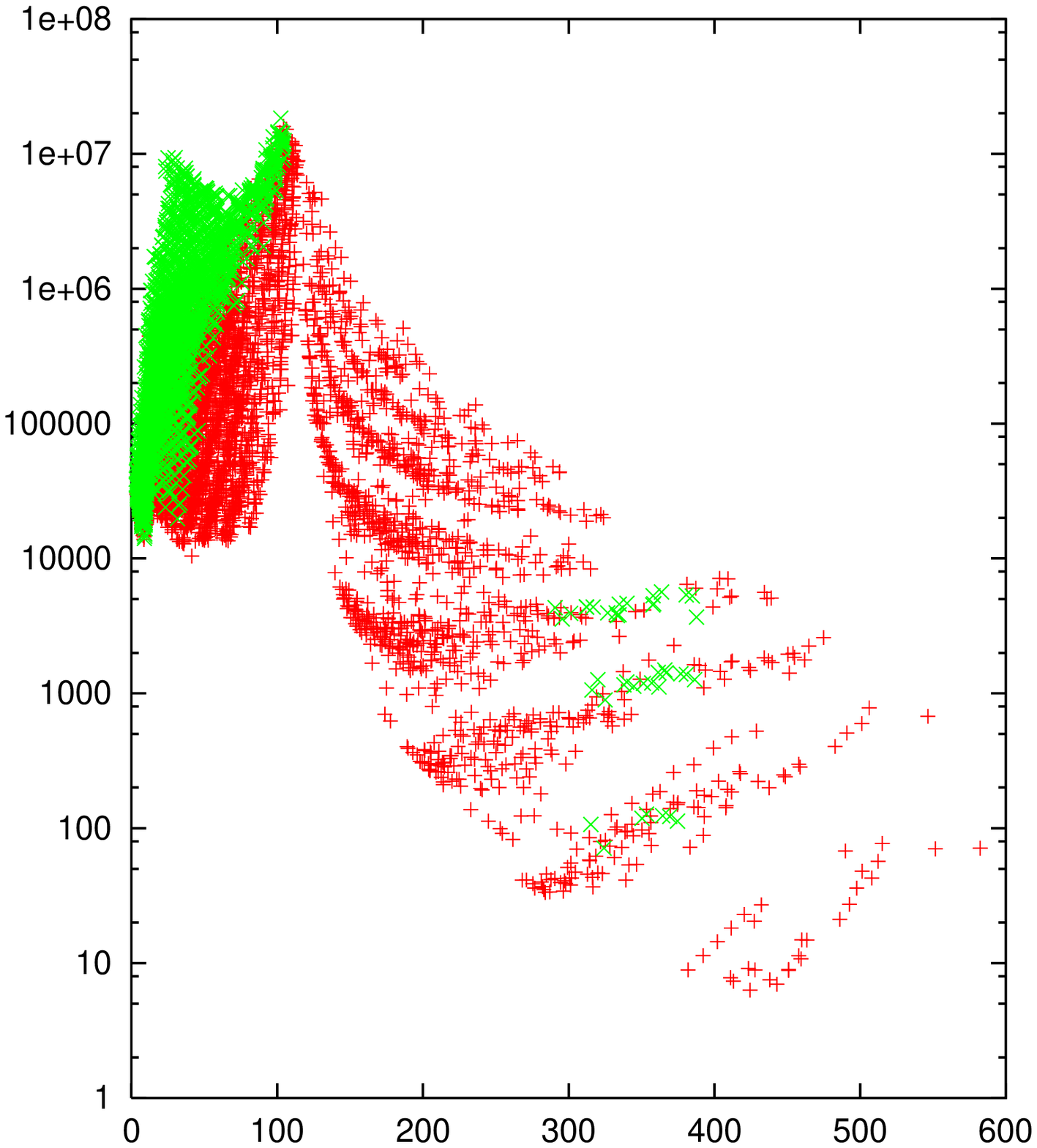,angle=0,height=5cm,width=\linewidth}
\end{minipage}\hfill\hfill
\begin{minipage}[b]{.495\linewidth}
\hskip1.25cm{\small Total number of $S_i$ events, $N_{S_i}$}
\centering\epsfig{file=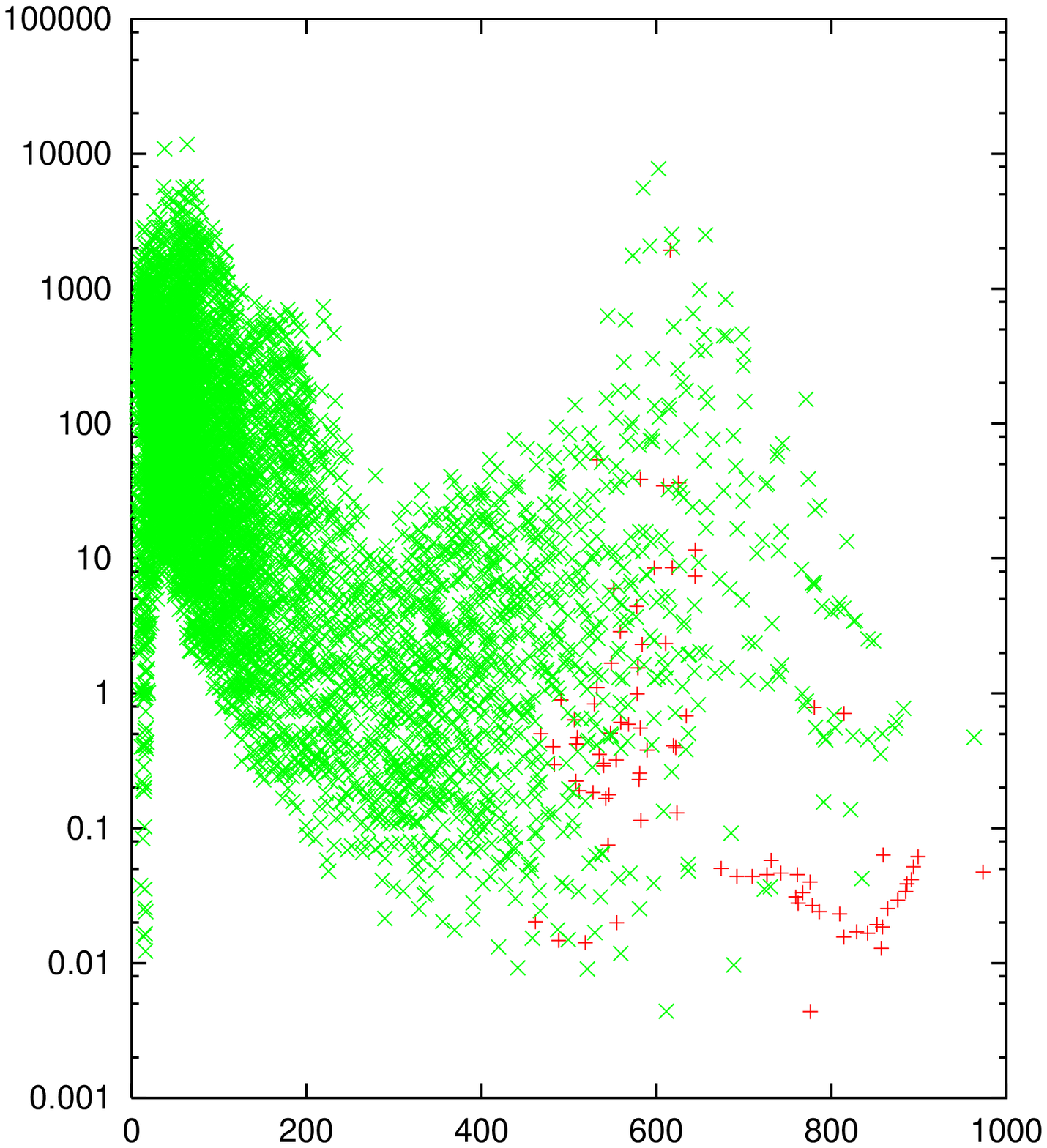,angle=0,height=5cm,width=\linewidth}
\end{minipage}\hfill\hfill
\begin{minipage}[b]{.495\linewidth}
\centering\epsfig{file=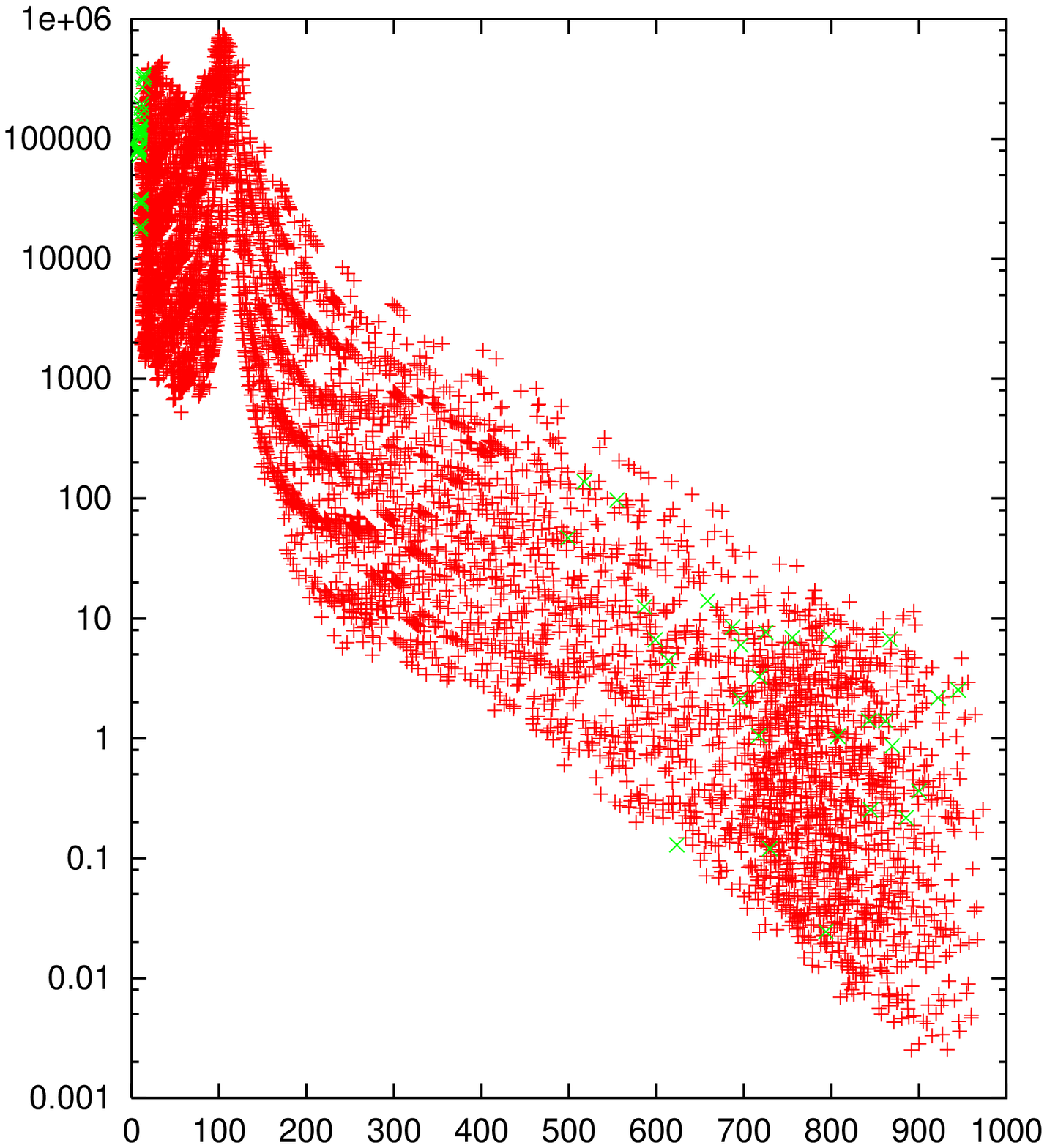,angle=0,height=5cm,width=\linewidth}
\end{minipage}\hfill\hfill
\begin{minipage}[b]{.495\linewidth}
\centering\epsfig{file=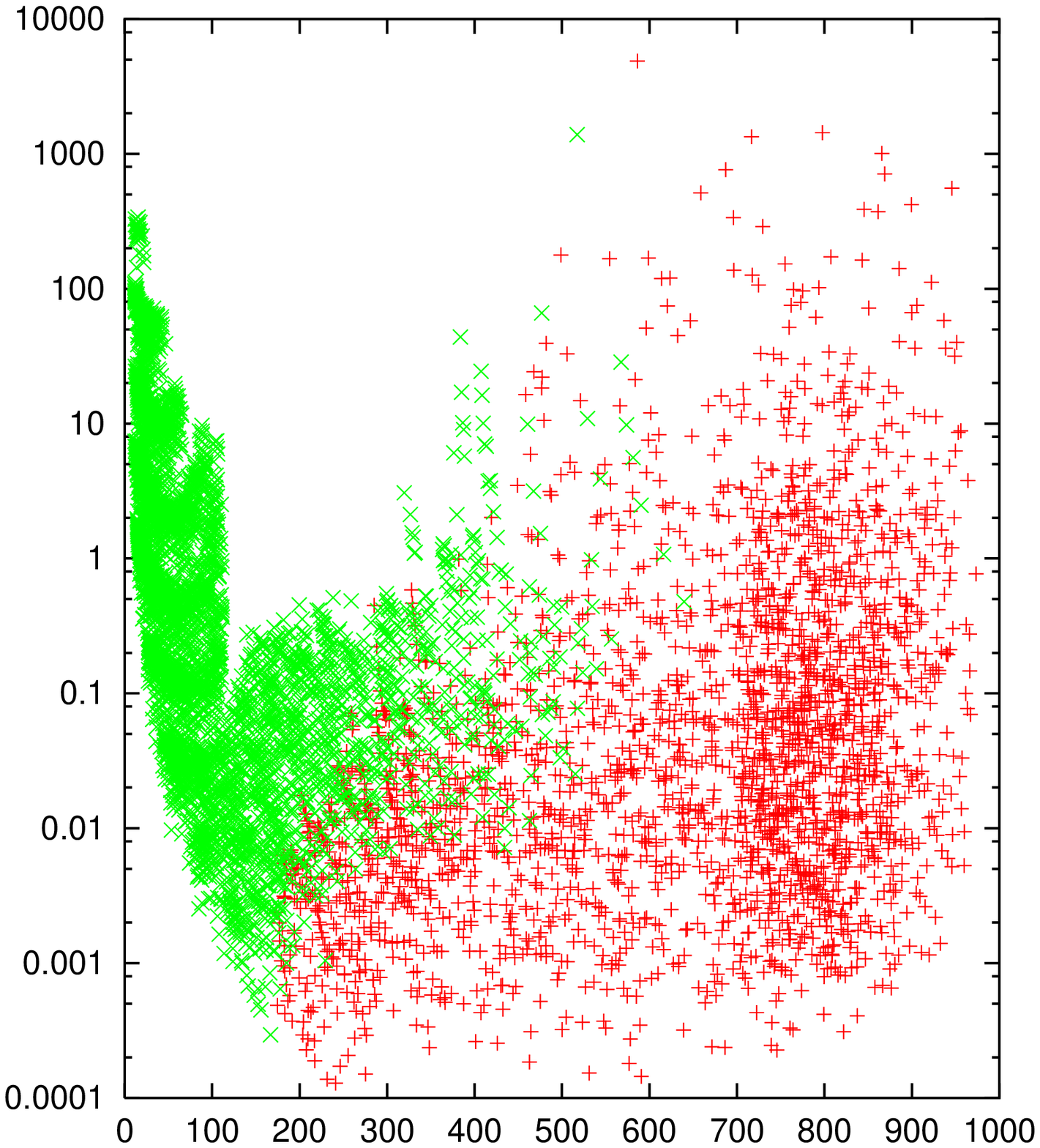,angle=0,height=5cm,width=\linewidth}
\end{minipage}

\hskip4.0cm{\small $M_{S_r}$ (GeV)}\hskip6.75cm{\small $M_{S_i}$ (GeV)}
\caption{As Fig.~\ref{fig:Tevatron} for the CP-even singlet $S_r$ (left plots) and the CP-odd singlet $S_i$
(right plots), at LHC after 300 fb$^{-1}$.}
\vspace*{-3mm}
\label{fig:LHC}
\end{figure}

For each of the five neutral Higgs bosons, we have computed the total number of
events obtained by summing the rates of all production processes in
(\ref{procs}), assuming 15 fb$^{-1}$ for Tevatron (Run II) and 300 fb$^{-1}$
for the  LHC, as integrated luminosities. We have plotted these rates versus
the mass of the given Higgs states in Figs.~\ref{fig:Tevatron}--\ref{fig:LHC}.
If, as threshold of detectability of a signal, we assume 100 events, the
conclusions are similar in both  models.

First, at Tevatron the non-singlet CP-even Higgs with mass 115 GeV, the qp
$H_u$, is of course always visible, but the other non-singlet CP-even state,
the qp $H_d$, is not, as it is too heavy. The non-singlet CP-odd state, the qp
$A$, will be visible if it is light enough ($M_A < 300-400$ GeV for a total
number of events $N_A > 100$). The singlet CP-even state, the qp $S_r$, will
also be visible up to masses of $\sim 300$ GeV if $\lambda \OOrd 10^{-3}$ (so
that this state is not too decoupled), particularly when it is quasi mass
degenerate with the qp $H_u$ state
(notice the peaks at $M_{S_r} \sim$ 115 GeV).
On the other hand, the singlet CP-odd
state, the qp $S_i$, will remain invisible at Tevatron. To render this
manifest, we have plotted in Fig.~\ref{fig:Tevatron} the total number of events
produced at Tevatron with $S_r$ in the final state, $N_{S_r}$ (two left plots)
in green (light) when the corresponding $A$ state is also visible ($N_A>100$)
and in red (dark) when it is not ($N_A<100$). Similarly, we did for $A$ (two
right plots), with green (light) when the corresponding $S_r$ is visible
($N_{S_r}>100$) and red (dark) when it is not ($N_{S_r}<100$). The upper plots
correspond to the NMSSM and the lower ones to the nMSSM. From these plots it is
easy to see in which cases one, two or three Higgs states will be visible (with
more than 100 produced events) at Tevatron.

At LHC, on the other hand, due to the large center of mass energy available,
all three non-singlet states, $H_u$, $H_d$ and $A$, will be visible. In the
singlet sector, the $S_r$ should be visible if its mass is $\Ord 600$ GeV and
$\lambda$ is not too small. In the NMSSM, this covers most of the parameter
space. Moreover, the CP-odd singlet, $S_i$ should be visible for an appreciable
part of the parameter space, at least for the NMSSM. These results are shown in
Fig.~\ref{fig:LHC}, with the same lay out and colour coding as for
Fig.~\ref{fig:Tevatron}, but this time for $S_r$ and $S_i$ at LHC.

Finally, 
it should be noted that the discovery areas of multiple Higgs boson states
identified in Figs.~\ref{fig:Tevatron}--\ref{fig:LHC} are indeed associated
to the same regions of parameter space. 
However, a first glance at the total number of CP-odd non-singlet $A$ 
produced at Tevatron in both models (Fig.~\ref{fig:Tevatron}, right-hand 
plots) 
might indicate that nearly all the parameter space of the models is already
covered by the CP-even singlet $S_r$ search at Tevatron, as all the plotted
points are in green (light). This is however not the case, as one can check
from the left-hand plots ($N_{S_r}$ vs. $M_{S_r}$ at Tevatron), where a lot of
points are still under the 100 events threshold. The fact that one sees only
green (light) points on the right-hand plots is due to the very high density of
points plotted, green (light) points being plotted after red (dark) ones.
Hence, there are red (dark) areas, uncovered by the $S_r$ searches, behind
green (light), covered, ones. This remark applies also for associated $S_r$,
$S_i$ searches at LHC (Fig.~\ref{fig:LHC}).
 
\section{Conclusions and final remarks}\label{summary}

The conclusions of this preliminary study are that, although the singlet sector
of non-minimal models tends to decouple from the rest of the spectrum in
the universal case, quasi pure singlet states could still be found at future
hadron colliders. One has to remember that a very light CP-even Higgs state is
not excluded by LEP searches if its coupling to gauge bosons is small enough.
Such a Higgs state should be looked for at Tevatron (Run II) where up to three
neutral Higgses could be found in our models (the CP-even non-singlet $H_u$ with
mass 115 GeV, and the CP-odd non-singlet $A$ and CP-even singlet $S_r$ if their
mass is small enough). On the other hand, the large center of mass energy of LHC
will allow us to see at least the three non-singlet Higgses ($H_u$, $H_d$ and
$A$) and its huge luminosity will trace the CP-even singlet state up to masses
of $\Ord 600$ GeV and, in some cases, the CP-odd one, making the whole neutral
Higgs spectrum visible.

The caveat of our analysis is that we have not performed a
full Higgs decay analysis in the NMSSM/nMSSM. 
One may question whether the additional Higgs states would actually 
be visible. For example, they would certainly couple to 
singlinos -- $\tilde S$ is always the Lightest Supersymmetric Particles (LSP)
in our context -- hence decay into the latter and thus remain undetected.
This should however not be the case. In fact, the coupling of the singlet 
states to ordinary matter are generally stronger in comparison
 (of order $\lambda$, whereas those to two 
singlinos are $\sim\lambda^3$). So that, in the end, the main decay channels
of singlet Higgs states  should be those into detectable 
fermions and gauge bosons.

\begin{acknowledgments}
C.H. would like to thank the late theory group of the Rutherford
Appleton Laboratory and the Theoretical Physics Department of Oxford, where
this work was achieved, for their kind hospitality. We are both grateful to 
the UK-PPARC for financial support.
\end{acknowledgments}


\end{document}